\documentclass[12pt]{spieman}  
\usepackage{amsmath,amsfonts,amssymb}
\usepackage{graphicx}
\usepackage{setspace}
\usepackage{tocloft}
\usepackage{multicol}

\title{PhoSim-NIRCam: Photon-by-photon image simulations of the James Webb Space Telescope's Near-Infrared Camera}

\author[a,b,*]{Colin J. Burke}
\author[a]{John R. Peterson}
\author[c]{Eiichi Egami}
\author[c]{Jarron M. Leisenring}
\author[a]{Glenn H. Sembroski}
\author[c]{Marcia J. Rieke}
\affil[a]{Purdue University, Department of Physics and Astronomy, 525 Northwestern Avenue, West Lafayette, IN, USA, 47907}
\affil[b]{University of Illinois at Urbana-Champaign, Department of Astronomy, 1002 W Green Street, Urbana, IL, USA, 61801}
\affil[c]{University of Arizona, Steward Observatory, 933 North Cherry Avenue, Tucson, AZ, USA, 85721}

\cftpagenumbersoff{figure}
\cftpagenumbersoff{table}
\begin{document}
\maketitle

\begin{center}
\textit{\footnotesize Draft version \today. Accepted for publication to SPIE JATIS.}
\end{center}

\begin{abstract}
\footnotesize Recent instrumentation projects have allocated resources to develop codes for simulating astronomical images. Novel physics-based models are essential for understanding telescope, instrument, and environmental systematics in observations. A deep understanding of these systematics is especially important in the context of weak gravitational lensing, galaxy morphology, and other sensitive measurements. In this work, we present an adaptation of a physics-based ab initio image simulator: The Photon Simulator (PhoSim). We modify PhoSim for use with the Near-Infrared Camera (NIRCam) --- the primary imaging instrument aboard the James Webb Space Telescope (JWST). This photon Monte Carlo code replicates the observational catalog, telescope and camera optics, detector physics, and readout modes/electronics. Importantly, PhoSim-NIRCam simulates both geometric aberration and diffraction across the field of view. Full field- and wavelength-dependent point spread functions are presented. Simulated images of an extragalactic field are presented. Extensive validation is planned during in-orbit commissioning. \end{abstract}


{\noindent \footnotesize\textbf{*}Colin J. Burke,  \linkable{colinjb2@illinois.edu} }

\begin{spacing}{1}   

\section{Introduction}
\label{sect:intro}  

\small

Development of high-fidelity image simulators has become commonplace in large instrumentation projects in astronomy~\cite{Peterson:2014,Peterson:2015,Leschinski2016,Rowe2015,Sivaramakrishnan2002,Dobke2010,Hilbert2017,Knight2012}. In addition, a full comprehensive physics-based method capable of simulating images from the source to the readout has been developed.\cite{Peterson:2014,Peterson:2015}. For example, galaxy morphology is altered by the atmosphere (for ground-based observatories), geometric aberrations in the optical train, diffraction, mirror micro-roughness, surface misalignments/perturbations, figure errors, and a variety of detector effects. Such systematics have important effects on morphological studies of astronomical objects. A good example is a weak lensing measurements, since systematically-induced galaxy ellipticity can contaminate shear measurements. Future and current dark matter surveys utilizing weak lensing as a probe of cosmological density fluctuations are limited by systematics\cite{Hikage:2018,Zhan2018,DES2018,Okabe:2010}. As telescopes become larger and their instruments become more sensitive, large extragalactic surveys will produce unprecedented levels of image data throughout the 2020s. This sharp rise in the available statistics must be informed by an equally transformative understanding of the systematics in these images.

The existing image simulation paradigm is mainly to use a parameterized point spread function (PSF) and create an image using the instrument's plate scale and background noise statistics. The approach of using optical path difference maps generated from a wavefront error budget is even rarely done in itself. However, tools such as WebbPSF\cite{Perren:2012,Perren:2014} employ this method for simulating images. WebbPSF relies on calculating the PSF from libraries of optical path difference maps. See Table~\ref{table:1} for a brief summary of PhoSim versus WebbPSF capabilities.

In this work, we present PhoSim-NIRCam: a comprehensive, end-to-end, image simulator of the James Webb Space Telescope's (JWST) Near-Infrared Camera (NIRCam) using a physics-based photon Monte Carlo code. This code, The Photon Simulator\cite{Peterson:2015,Peterson:2014} (PhoSim), enables detailed study of the optical system and detector effects including the field- and wavelength-dependent PSF. Forward-modeling approaches such as those presented in this work are still rarely employed for astronomical image simulations. In this work, we study telescope/instrument systematics in images produced by PhoSim-NIRCam, which simulates one photon at a time. Notably, PhoSim-NIRCam can simulate both the diffraction and geometric aberration components of the PSF across the field of view. Modular PhoSim commands can be used with ease to turn various physics on and off.

Additionally, we report on changes made to the PhoSim source code by the authors of this paper to simulate infrared space-based telescopes. As of version 4.0, the NIRCam instrument's imaging modes are fully implemented in PhoSim. The PhoSim code is publicly available and open-source\footnote{\footnotesize\url{https://bitbucket.org/phosim/phosim_release}}.

\subsection{JWST/NIRCam}

JWST~\cite{Gardner:2006} is NASA's next-generation flagship space-based observatory. JWST will be located at Earth-Sun Lagrange point 2, and is currently slated for launch in 2021. The observatory has a planned minimum mission lifetime of five years with a commissioning phase of six months. Its primary imaging instrument, NIRCam~\cite{Greene:2010} is a dual-channel optical system designed to operate in the wavelength range at 0.6 $\mu$m to 5.0 $\mu$m. NIRCam has a broad range of scientific goals including subjects where telescope/instrument systematics are important. See the list of approved guaranteed-time observers (GTO) and director's discretionary early release science (ERS) programs for specific examples.

\subsection{The Photon Simulator}

PhoSim is a set of physics-based, end-to-end, ab initio photon Monte Carlo codes originally developed for use with the Large Synoptic Survey Telescope\cite{LSST2009,Ivezic:2008} (LSST).

While LSST is a very different telescope than JWST, the PhoSim architecture is generalized in a way that makes implementing new telescopes straightforward. We simply add a new set of instrument and site characteristics (ISC) data files that specify the details of JWST/NIRCam and its environment. This allows us to generate high-fidelity NIRCam images quickly while taking advantage of PhoSim's extensive validation and robustness obtained over its more than 10 year development period.

One important benefit of PhoSim is its speed. With multithreading capability, PhoSim-NIRCam can produce images from a moderately sized catalog on a modern laptop or desktop computer in just a few minutes, whereas a PSF from a single faint star can be simulated in milliseconds.

\begin{figure}[ht]
\centering
\includegraphics[width=0.75\columnwidth]{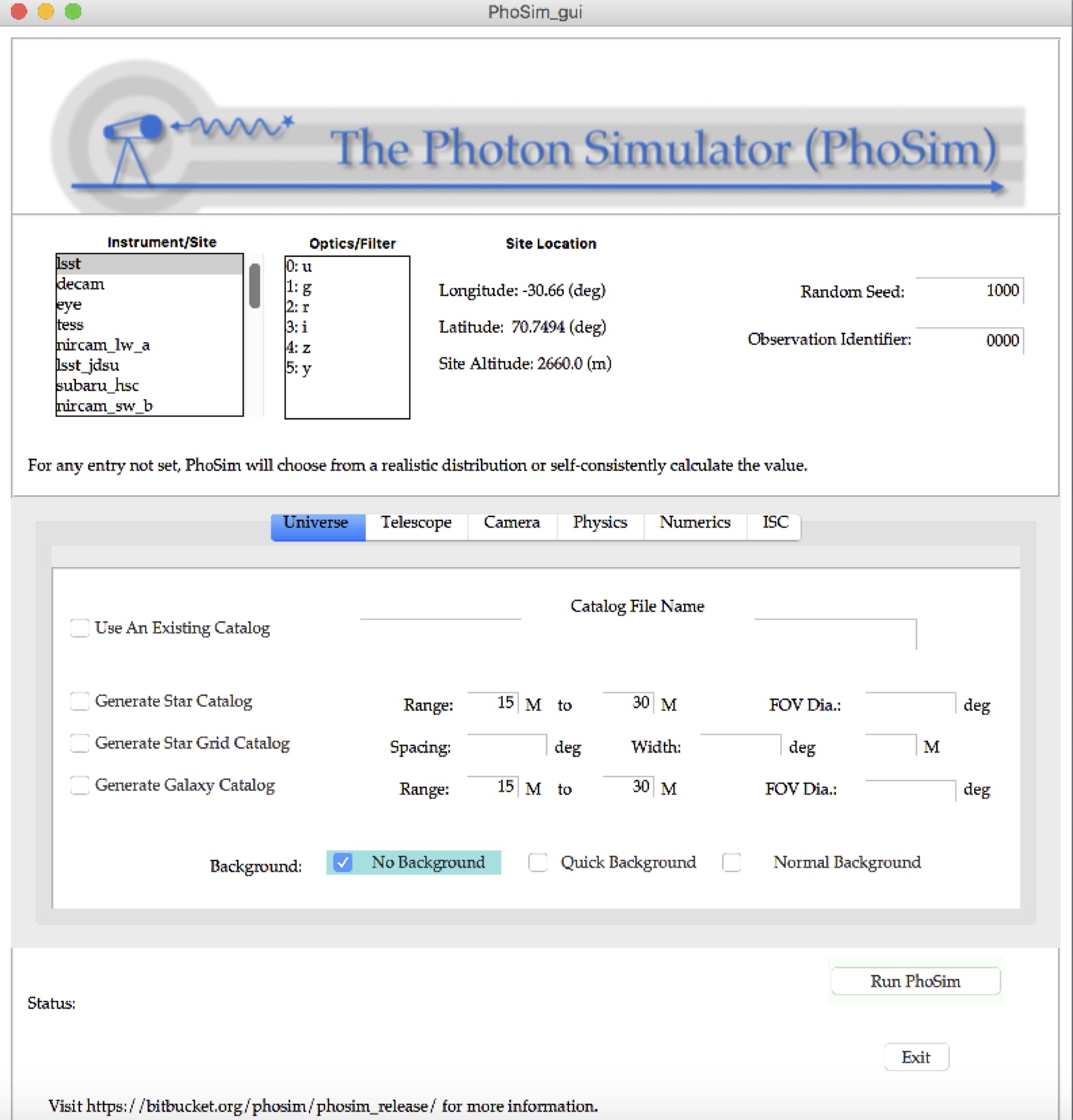}
\caption{\label{fig:gui}
Screenshot of the PhoSim version 4.0 GUI running on MacOS. The telescope/instrument must be specified. All other inputs are optional; PhoSim will choose settings from distributions or calculate them self-consistently.
}
\end{figure}

Additionally, PhoSim can be run on a single laptop with a modern graphical user interface (GUI; Fig.~\ref{fig:gui}) or command line, while also being scalable to grid or cluster computing. Large data challenges are already underway for LSST and its survey to test its image processing pipeline.

Next, the physics-based nature of PhoSim-NIRCam calculations means that complicated effects in images naturally emerge from the underlying physics once the proper ISC files are provided. In its over a decade of development as the official image simulator for the LSST project, PhoSim's physics have been successfully validated against its expected behavior from optical models such as Zemax.

Perhaps the most powerful component of PhoSim is the option to independently turn on and off various components of its physics. Using simple, one-line commands, one can determine how each component of the physics in PhoSim affects the final images. In this work, for example, images are simulated with and without diffraction to investigate NIRCam PSFs.

PhoSim works by performing a comprehensive photon-by-photon Monte Carlo raytrace through the atmosphere (which we turn off using a physics command), telescope, camera, and detector. The effects of diffraction are included by performing Fourier transform of the JWST entrance pupil and kicking each photon's angle proportionally. The result is an output of Flexible Image Transport System (FITS) images after the raytrace (electron image) and after the readout procedure (amplifier image). The electron image essentially provides a pre-calibrated image, whereas the amplifier images replicates the noise seen in raw data.

\begin{table}[]
\small
\centering
\begin{tabular}{l|ll}
 & WebbPSF & PhoSim-NIRCam \\ [0.5ex] \hline
Optics: & Library of OPD maps & Full 3-D optical model \& perturbation capability \\
Simulates: & PSF only & PSF and full image \\
Modes: & Imaging and chronograph modes & Imaging modes only \\
Detector: & No detector model & HgCdTe detector and limited noise model \\
Interface: & GUI or python API & GUI or command-line interface \\ [1ex]
\end{tabular}
\caption{Summary of PhoSim-NIRCam versus WebbPSF capabilities.}
\label{table:1}
\end{table}

\section{Implementation}

In this work, we report on the implementation of several new PhoSim features in version 4.0 to simulate infrared and space-based telescopes: proper treatment of the pupil diffraction, Mercury Cadmium Telluride (MCT) detectors, and a mode for MULTIACCUM readout patterns.

NIRCam is a dual-channel optical system for short wavelength (SW) and long wavelength (LW) infrared (IR) light bifurcated with a dichroic beam splitter. Two ``fully redundant'' and ``functionally identical'' modules, denoted A and B, contain both channels --- meaning four focal plane assemblies in total\cite{Greene:2010}. However, there are several key differences between the modules such as the throughput curves and detector parameters. Thus, we create four separate ISC file sets for each NIRCam channel/module: \verb$nircam_sw_a$, \verb$nircam_sw_b$, \verb$nircam_lw_a$, \verb$nircam_lw_b$.

\section{Methodology}

Setting aside additional effects, the total PSF for space telescopes is comprised of a geometric component (from aberrations in the optical train) and a diffraction component (from the limiting pupil geometry). In the diffraction-limited regime the characteristic size of the geometric component of the PSF is negligibly small compared to the diffraction-limit. However, in many cases the geometric component of the PSF is non-negligible, such as when the image is out of focus or when aberrations are large at certain wavelengths or field positions. When no atmosphere is considered, the condition for this intermediate case occurs when the characteristic size of the geometric and diffraction components of the PSF are roughly equal,
\begin{equation}
\sigma_g \sim \frac{\lambda F}{D}
\end{equation}

where $\lambda$ is the photon wavelength, $F$ is the effective focal length, $D$ is the effective pupil diameter, and $\sigma_g$ is characteristic size of the geometric component of the PSF.

We can rewrite the above expression as,
\begin{equation}
\frac{\lambda f}{\sigma_g} \sim 1
\end{equation}
upon substituting for the focal ratio $f=F/D$. For an Airy-like PSF, this condition is: $\sigma_g \sim 0.42 \lambda f$, or in terms of the diffraction-limited angular size: $\theta \sim 1.22 \lambda / D$.

We are interested in the subtleties of this regime. Additionally, NIRCam is only required to be diffraction-limited for photon wavelength's above 2 $\mu$m. Thus, a comprehensive approach that accounts for both geometric and diffraction components is required to capture the full PSF morphology below 2 $\mu$m. To accomplish this, a complete physical description of JWST/NIRCam is implemented for each component of the PSF. When simulating the geometric component of the PSF, PhoSim uses a Monte Carlo raytrace method through a comprehensive specification of the optical prescription. When simulating the diffraction component of the PSF, we use the standard results in the Fraunhofer regime where the diffraction component of the PSF is given by the Fourier transform of the limiting pupil geometry.

Using its powerful physics commands, PhoSim has the capability to simulate any and all components of the PSF independently or together in various combinations. The following subsections detail our methodology for reproducing the major components of the PSF.

\subsection{Geometric Optics}

The NIRCam optical design is complicated due to the unique constraints of a space-based observatory. There are many flat fold mirrors that reflect light back and forth through a series of lenses that achieves the long focal length in a compact design. This means that there are a large number of optical surfaces (28 for the SW channel; 26 for the LW channel) with various orientations and positions. The design is also slightly off-axis by about $0.13$ degrees. Since PhoSim is physics-based, correctly implementing the optical design is essential to producing realistic images with the expected PSF and field distortion.

\begin{figure}
\centering
\includegraphics[width=1.0\columnwidth]{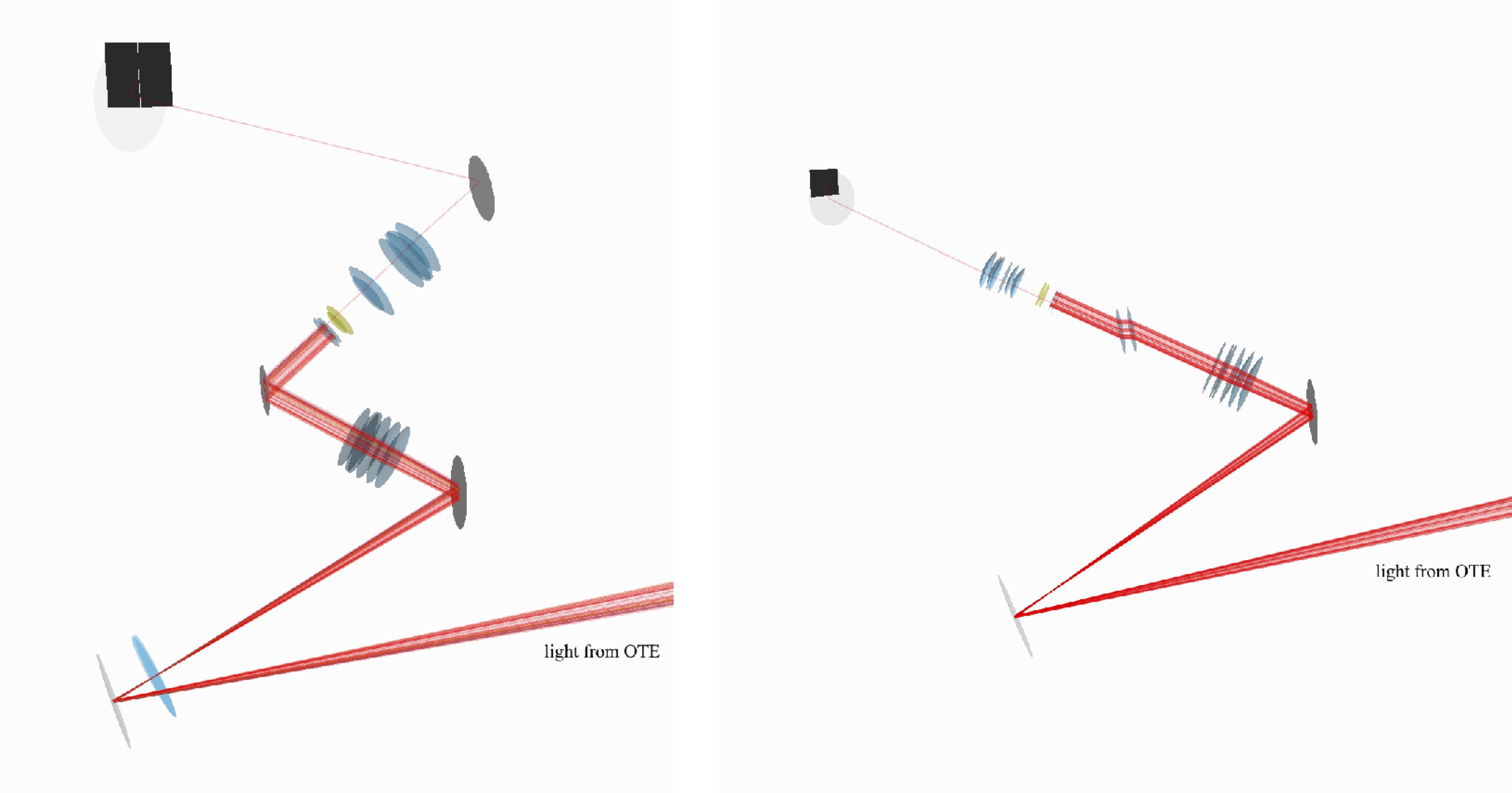}
\\
(a) \hspace{6.5cm} (b)
\caption{\label{fig:nircam}
Visualization of a PhoSim Monte Carlo photon raytrace through NIRCam SW (a) and LW (b) channels. Mirrors are shown in gray, lenses in blue, filters in yellow, and detector chips in black. Only a sample of rays are shown to avoid clutter. As a result, few photons are transmitted through the filters. The raytrace from previous OTE surfaces is also included, but not shown here.}
\end{figure}

The PhoSim optics model is a full description of the optical prescription of JWST/NIRCam converted from two Lockheed Martin flight-ready Zemax lens files: \verb$L050713FLT.zmx$ (LW channel) and \verb$S050713FLT.zmx$ (SW channel). The Zemax lens files contain a complete description of the optical system where spatial coordinates of each surface are defined sequentially \cite{Zemax:2011}. These models approximate OTE primary mirror as a single cylindrically-symmetric surface.

The JWST optical train can be divided into two components: the optical telescope element (OTE) and integrated science instrument module (ISIM). The OTE is a three-mirror anastigmat design comprised of the segmented primary mirror plus three additional surfaces (including the planar fine steering mirror). The ISIM houses all four of the observatory's instruments and is cryogenically cooled to 37 K to reduce thermal noise. Together, the entire optical train is referred to as the optical telescope element/integrated science instrument module (OTIS). OTIS testing was completed successfully in 2017\cite{Kimble2018}.

In this work, we implement the entire OTIS optical prescription from the flight-ready Zemax design files which exist for both modules (Fig.~\ref{fig:nircam}). Each optical surface has parameters describing its coordinates and shape. The coordinate parameters are 3 Euler angle orientations and 3 spatial vertex positions. The shape parameters are contained in the typical equation for cylindrically symmetric surface,
\begin{equation}
z(r)=\frac{r^2}{R\left(1+\sqrt[]{1-(1+\kappa)\frac{r^2}{R^2}}\right)}+\sum_{i=3}^{10}\alpha_ir^i
\end{equation}
where the sag of the surface $z$ is expressed in terms of the radial distance from the optical axis $r$, the radius of curvature $R$, the conic constant $\kappa$, and the asphere coefficients $\alpha_i$. The the inner and outer radius is specified for each surface. The finite elements of the surface map are far smaller than the wavelength of optical and infrared light. The primary mirror is specified in this manner where the inner and outer radii have been calibrated to reproduce the expected wavefront error and photometry.

In addition, we include models for the dispersive index of refraction of intervening medium from Zemax. Five materials are modeled in PhoSim for NIRCam: BaF$_2$, LiF$_2$, ZnSe, Si, and fused silica. The cryogenic indicies of refraction for each material are described as a function of wavelength $\lambda$ by either the Sellmeier equation \cite{Sellmeier:1871}:
\begin{equation}
\begin{aligned}
n(\lambda)=\sqrt{1+\frac{B_1\lambda^2}{\lambda^2-C_1}+\frac{B_2\lambda^2}{\lambda^2-C_2}+\frac{B_3\lambda^2}{\lambda^2-C_3}}
\end{aligned}
\end{equation}
or the Schott equation \cite{Zemax:2011}:
\begin{equation}
\begin{aligned}
n(\lambda)=\sqrt{a_o+a_1\lambda^2+a_2\lambda^{-2}+a_3\lambda^{-4}+a_4\lambda^{-6}+a_5\lambda^{-8}}.
\end{aligned}
\end{equation}
where the various coefficients are specified for a material.

The throughput curves (reflection, transmission, and absorption probability as a function of wavelength and incident angle) of the entire optical system are taken from the JWST user documentation\cite{STScI2017}. We divide-out the detector quantum efficiency, which is calculated with electron conversion physics with PhoSim. The desired NIRCam channel, module, and filter can be selected from the GUI or by specifying the appropriate command-line inputs to PhoSim.

\subsection{Diffraction}

The diffraction component of the PSF is given by the squared modulus of the Fourier transform of the electric field amplitude over the pupil plane,
\begin{equation}
\text{PSF}(\hat{n})=\left|\frac{1}{A}\int_Ae^{i\hat{n}\cdot \vec{r}} e^{i\phi(\vec{r})}d^2\vec{r}\right|^2
\end{equation}
where the wavenumber is $k=2\pi/\lambda$, $\hat{n}$ is the unit vector in the direction of a field point on the focal plane, $\phi$ is the phase shift induced by the changing index of refraction along the optical  train, and $A$ is the pupil screen function transmission probability.

\begin{figure}[ht]
\centering
\includegraphics[width=0.75\columnwidth]{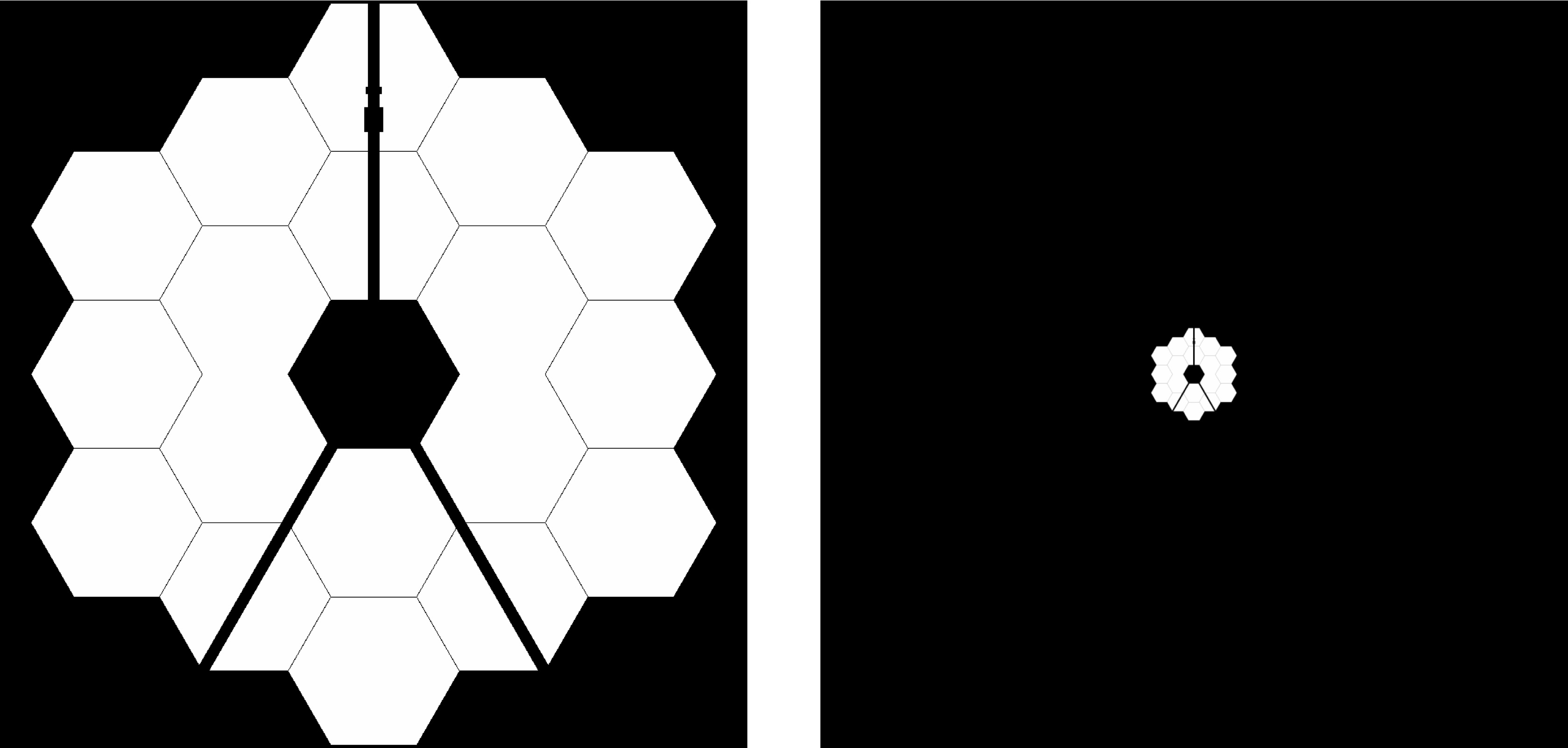}
\\
(a) \hspace{5.7cm} (b)
\caption{\label{fig:pupil}
JWST revision V pupil screen before zero-padding (a) and after (b).}
\end{figure}

\begin{figure}[ht]
\centering
\includegraphics[width=0.5\columnwidth]{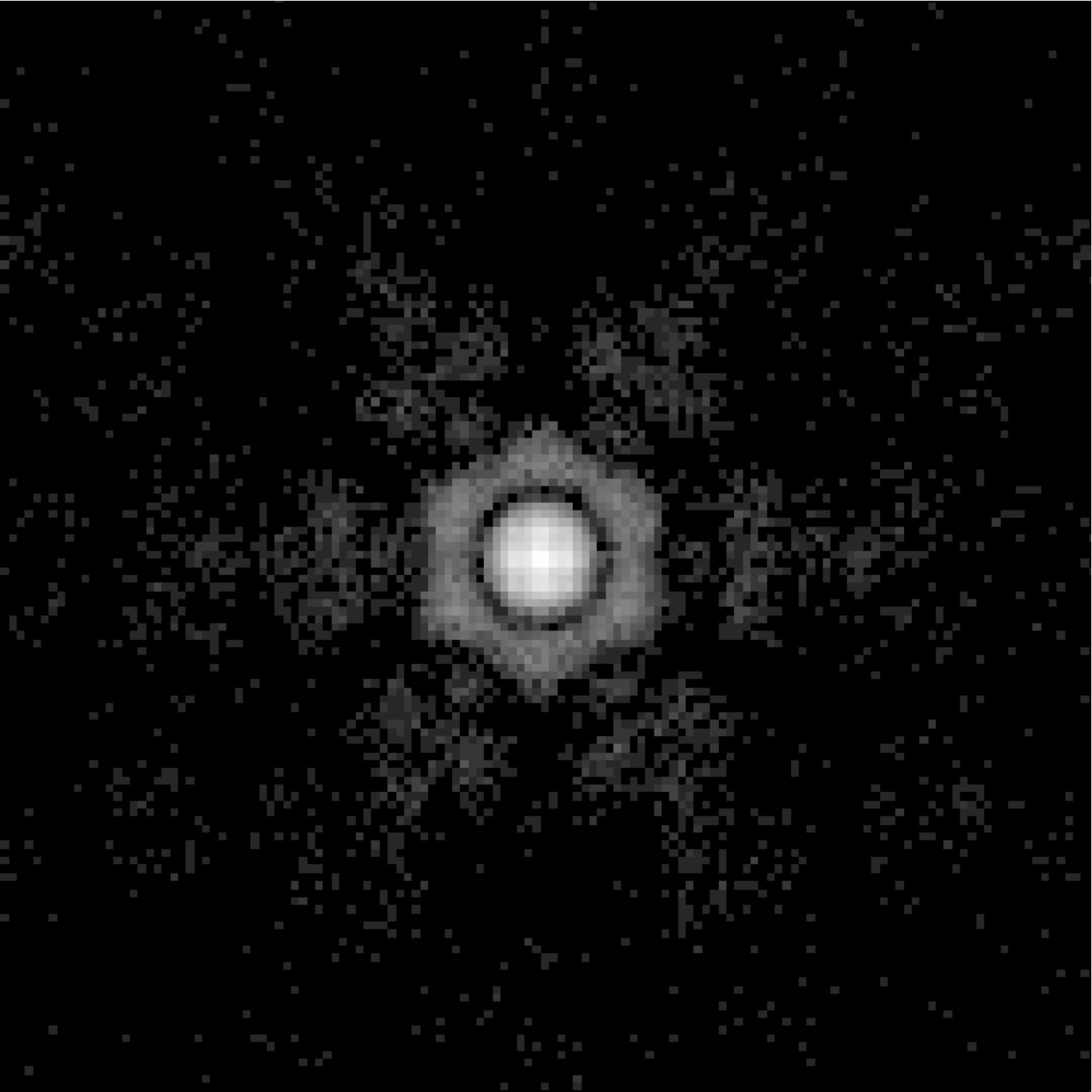}
\caption{\label{fig:diffEx}
An oversampled example of a PhoSim-NIRCam PSF with only diffraction physics on.}
\end{figure}

When simulating the diffraction component of the PSF, the basis of the algorithm is a fast Fourier transform (FFT) of the 2-D pupil geometry. Due to the nature of the discrete FFT algorithm, we must pad the array with zeros to ensure enough frequency bins are created to obtain reasonably accurate results (Fig.~\ref{fig:pupil}). Then, we create a cumulative probability distribution,
\begin{equation}
P(\vec{r}) = \int_A \text{FFT}(A(\vec{r})) d\vec{r}
\end{equation}
The result is to kick photons' incident angle $\theta$ before the raytrace through the optics by,
\begin{equation}
\delta\theta=\frac{\left|\vec{r}\right|\lambda}{\gamma}
\end{equation}

where $\vec{r}$ is sampled from a uniformly distributed random number in $P(\vec{r})$, $\lambda$ is the wavelength of the photon, and $\gamma$ is the zero-padding factor. Due to the extra zero-padding, we lose some detail in the pupil image since it must be scaled down to fit inside an array of reasonable size ($1024 \times 1024$ pixels). However, our analysis to follow demonstrates our approximation is reasonably good at replicating the expected diffraction pattern and PSF size with a padding factor of $\gamma=8$ (Fig.~\ref{fig:diffEx} and \ref{fig:psfsize3}).

Presently, the OTE is described by a single cylindrically-symmetric surface for the geometric raytrace, and a tricontagon-shaped pupil screen (Fig.~\ref{fig:pupil}) which is FFT’d separately. Photons are then kicked by an angle $\delta\theta$ (from result of FFT) and rays are propagated geometrically to the focal plane to simulate diffraction from the tricontacgon aperture. Although the higher-order spatial content of the geometric PSF is affected by the segmented primary mirror surface geometry, we show the convergence to the expected diffraction limit in Fig.~\ref{fig:psfsize1} and compare (to first-order) the PSF size to the nominal diffraction limit and WebbPSF (Fig.~\ref{fig:psfsize3}).

\subsection{Photo-Electric Conversion}

NIRCam's ten $2048 \times 2048$ pixel Teledyne HAWAII-2RG complementary metal-oxide-semiconductor (CMOS) detectors are composed of MCT, Hg$_{1-x}$Cd$_x$Te, with different relative compositions (molar fraction or stoichiometric ratio) of Cd to Hg $x$~\cite{Loose:2007}. This allows for a tunable bandgap, which corresponds to a variable cutoff wavelength $\lambda_{co}$. Considerable effort has been made to understand the optical properties and electron interactions of MCT photodetectors in recent decades~\cite{Rogalski:2005,Itsuno:2012,Rieke:2007}. We have implemented MCT detectors in PhoSim, which calculates the photon mean free path from the absorption coefficient for a given $x$.

PhoSim simulates all relevant physics of CMOS (and CCD) detectors in a multi-step photon-to-electron conversion code. A final image is produced with highly realistic results \cite{Peterson:2015}. However, in previous versions of PhoSim, only Si detectors were implemented. To model the absorption coefficient in MCT as a function of photon wavelength in the absorption layer, we first make use of the Hansen equation\cite{Hansen:1982} describing the material's energy gap:
\begin{equation}
\label{hanson}
\begin{aligned}
E_g(x,T)=-0.302+1.93x+5.53(10^{-4})T(1-2x)-0.810x^2+0.832x^3
\end{aligned}
\end{equation}
where $E_g$ is the bandgap energy in eV. Applying the Planck-Einstein relation, $E_g = hc/\lambda_{\text{cutoff}}$, Eq.~\ref{hanson} can be re-expressed in terms of the cutoff wavelength, given in $\mu m$:
\begin{equation}
\label{hanson_co}
\begin{aligned}
\frac{1.24\text{ eV$\mu$m}}{\lambda_{\text{cutoff}}}\cong-0.302+1.93x+5.53(10^{-4})T(1-2x)-0.810x^2+0.832x^3.
\end{aligned}
\end{equation}
Using the known cutoff wavelengths of both detectors, $\lambda_{\text{cutoff}}=2.5~\mu$m and $\lambda_{\text{cutoff}}=5.3~\mu$m \cite{Garnett:2004} for the SW and LW channels respectively, we solve for the real root of Eq.~\ref{hanson_co} with NIRCam's cryogenic temperature $T = 37$ K. The small effect on the absorption coefficient from variations of $x$ in the absorption layer is not currently considered.

To calculate the absorption coefficient $\alpha$, we implement an empirical piece-wise model for the Kane region ($E_{\gamma} > E_g$) given by Chu et al. \cite{Chu:1994} and the modified Urbach tail ($E_{\gamma} < E_g$), given by Finkman and Schacham \cite{Finkman:1984,Hougen:1989} where $E_{\gamma}$ is the incident photon energy:

\begin{equation} \alpha = \begin{cases}
\alpha_o\exp{\left[\sigma \left( \frac{E_\gamma-E_o}{T+T_o}\right)\right]} & E_\gamma < E_g\\
\beta \sqrt[]{E_\gamma-E_g} & E_\gamma > E_g
\end{cases}
\end{equation}

where the parameters are defined as:

\begin{multicols}{2}

$\alpha_o=\exp{(53.61x - 18.88)}$

$E_o = -0.3424 + 1.838x + 0.148x^2$

$T_o = 81.9$

$\sigma = 3.267\times10^{4}(1 + x)$
\\
$E_T = \left(\frac{T_o + T}{\sigma}\right)\ln(\alpha_T/\alpha_o) + E_o$

where $\alpha_T = 100 + 5000x$
\\
$\beta = \alpha_T(E_T - E_g)^{-1/2}$
\\
and $E_g$ is specified by Eq.~\ref{hanson}.

\end{multicols}

\begin{figure}[ht]
\centering
\includegraphics[width=0.7\columnwidth]{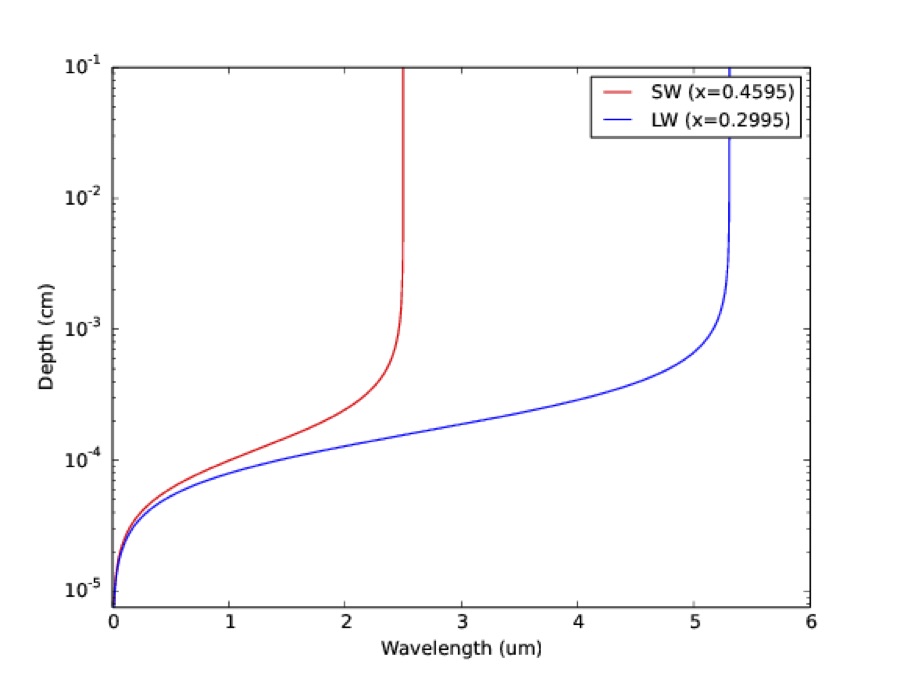}
\caption{
\label{fig:mfp}
Plot of the inverse of the absorption coefficient (mean free path) for both SW and LW MCT types as a function of wavelength. Results are consistent with Ref.~\citenum{Beletic:2008}, Fig. 5.
}
\end{figure}

The mean free path of a photon is simply given as the inverse of $\alpha$. The conversion path length is calculated in PhoSim by multiplying the absorption coefficient by an exponentially distributed random number~\cite{Peterson:2015}. Both detectors' absorption regions are approximately 8 $\mu$m thick. Fig.~\ref{fig:mfp} shows the absorption coefficients for MCT in PhoSim as a function of incident photon wavelength for both channels at 37 K.

Next, we implement a simple model given by Lui et al.\cite{Lui:1994} for the index of refraction in MCT as a function of $\lambda$, $T$, and $x$:
\begin{equation}
\begin{aligned}
n(\lambda,T,x)=\sqrt[]{A+\frac{B}{1-(C/\lambda)^2}+D\lambda^2}
\end{aligned}
\end{equation}
where the parameters A, B, C, and D are defined as:
\\

$A=13.173 - 9.852x + 2.909x^2 + 0.001(300 - T)$

$B=0.83 - 0.246x - 0.0961x^2 + 8\times 10^{-4}(300 - T)$

$C = 6.706 - 14.437x + 8.531x^2 + 7\times 10^{-4}(300 - T)$

$D = 1.953\times10^{-4} - 0.00128x + 1.853\times10^{-4}x^2$.
\\

In accordance with Ref.~\citenum{Dornhaus:1983}, the relative permittivity (dielectric constant) in MCT $\epsilon_\text{MCT}$ is given in the high frequency approximation by,
\begin{equation}
\begin{aligned}
\epsilon_\text{MCT}(x)=15.2-15.6x+8.2x^2. 
\end{aligned}
\end{equation}

The transverse diffusion is calculated with the Gaussian diffusion width, $\sqrt{2Dt_c}$, where $D$ is the diffusion coefficient given by,
\begin{equation}
\begin{aligned}
D=\frac{\mu_q(x,T)kT}{q}
\end{aligned}
\end{equation}
where $\mu_q(x,T)$ is the electron mobility in MCT. Following Ref.~\citenum{Rosbeck:1982}, we implement the model for electron mobility in MCT:
\begin{equation}
\begin{aligned}
\mu_q(x,T) = \frac{9\times10^8s}{100T^{2r}}
\end{aligned}
\end{equation}
where $r=(0.2/x)^{0.6}$ and $s=(0.2/x)^{7.5}$.
The collection time is,
\begin{equation}
\begin{aligned}
t_c=\int_{z_c}^{z}\frac{dz}{|\mu_q(x,T)E_z(z)|}.
\end{aligned}
\end{equation}

Further work will identify what other relevant sensor effects may be important to produce more realistic images.

\subsection{Device Readout}

We mimic the standard NIRCam CMOS readout procedure, dubbed MULTIACCUM. This means that multiple frames can be read out non-destructively during an integration sequence as charge accumulates in the pixels. In practice, multiple frames are average-combined into groups and an initial reference (zero) frame before they are reset for the next sequence due to data transmission constraints. Several different MULTIACCUM sequence patterns exist depending on the observation's science goals, target, and time constraints.

Each chip is segmented into four output channels of dimensions $2048\times512$ for the readout. Thus, four average-combined amplifier images are generated for each MULTIACCUM group plus the reference frame. The proper read noise and bias level is added (see Ref.~\citenum{Peterson:2015} for a more complete description of the PhoSim amplifier images.)

\subsection{Background}
The background in deep NIRCam images is dominated by Zodiacal light. There exists thermal emission from JWST itself, but this is negligible in the  NIRCam bands. The Zodiacal light spectral radiance is the sum of scattered and thermal emission produced by the Zodiacal dust. The model given by in the sensitivity calculations technical report \cite{Rieke:2013} is,

\begin{equation}
\begin{aligned}
F(\lambda)=\frac{3.95\times10^{-14}\cdot1.19\times10^8\cdot\lambda^{-5}}{e^{14388/(\lambda\cdot5300)}-1}+\frac{2.79\times10^{-8}\cdot1.19\times10^8\cdot\lambda^{-5}}{e^{14388/(\lambda\cdot282)}-1}
\end{aligned}
\end{equation}

where $\lambda$ is given in $\mu$m. The first term is the scattering and the second term is the thermal emission.

The spatial and temporal variation of the Zodiacal background flux is also modeled. To first-order, the spatial variation is a function of the ecliptic latitude of the telescope's pointing. The temporal variation is a seasonal variation with a period of 1 year\cite{Kelsall:1998}.

\begin{figure}[ht]
\centering
\includegraphics[width=0.9\columnwidth]{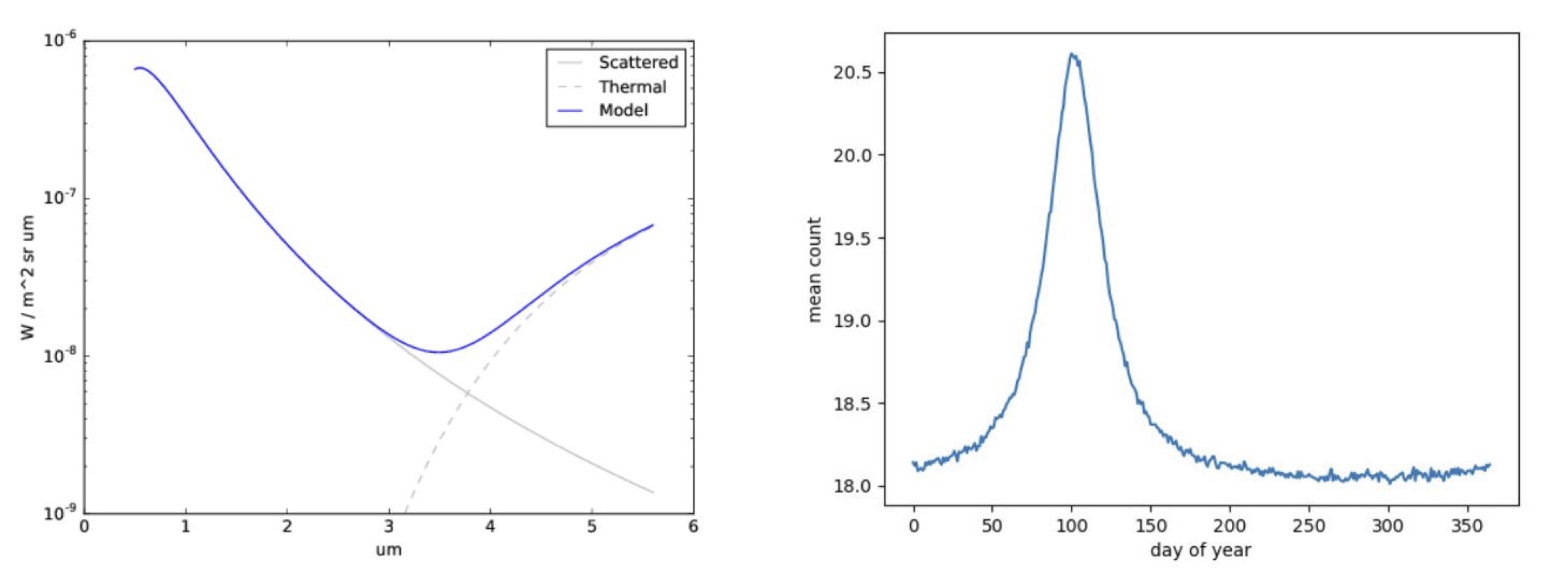}
\\
(a) \hspace{6.7cm} (b)
\caption{\label{fig:zod}
PhoSim Zodiacal background SED model (a) and Zodiacal background time-variation model for a fixed telescope pointing direction (b).
}
\end{figure}

Cosmic rays can be simulated in PhoSim by applying a known rate of comics ray events and randomly selecting from a set of pre-defined postage stamp images of cosmic ray interactions which are then given a random orientation. This is then simply added to the electron and amplifier image outputs.

\section{Results and Discussion}

\subsection{Point Spread Function}

Systematics effects in the PSF can be wavelength and field dependent. We determine the centroid RMS size of the PSF at various positions in the NIRCam field of view (field points) using each filter.

The total PSF size $\sigma_T$ can be approximated by the sum in quadrature of the geometric $\sigma_g$ and $\sigma_d$ diffraction components,
\begin{equation}
\label{eq:diff}
\sigma_T\approx\sqrt[]{\sigma_g^2+\sigma_d^2}.
\end{equation}
We adopt an iterative weighting method used in the weak lensing community where the RMS $\sigma$ is,
\begin{equation}
\sigma=\sqrt[]{I_{11}+I_{22}}
\end{equation}
where $I_{ij}$ are normalized moments of the PSF's intensity profile $f(x_1,x_2)$ weighted by an elliptical Gaussian filter $W(x_1,x_2)$,
\begin{equation}
I_{11}=\frac{\int\int dx_1dx_2W(x_1,x_2)f(x_1,x_2)x_ix_j}{\int\int dx_1dx_2W(x_1,x_2)f(x_1,x_2)}.
\end{equation}

This method is iterated until we reach an acceptable level of error, which is given in the image captions wherever measured values appear.

\begin{figure}[ht]
\centering
\includegraphics[width=0.7\columnwidth]{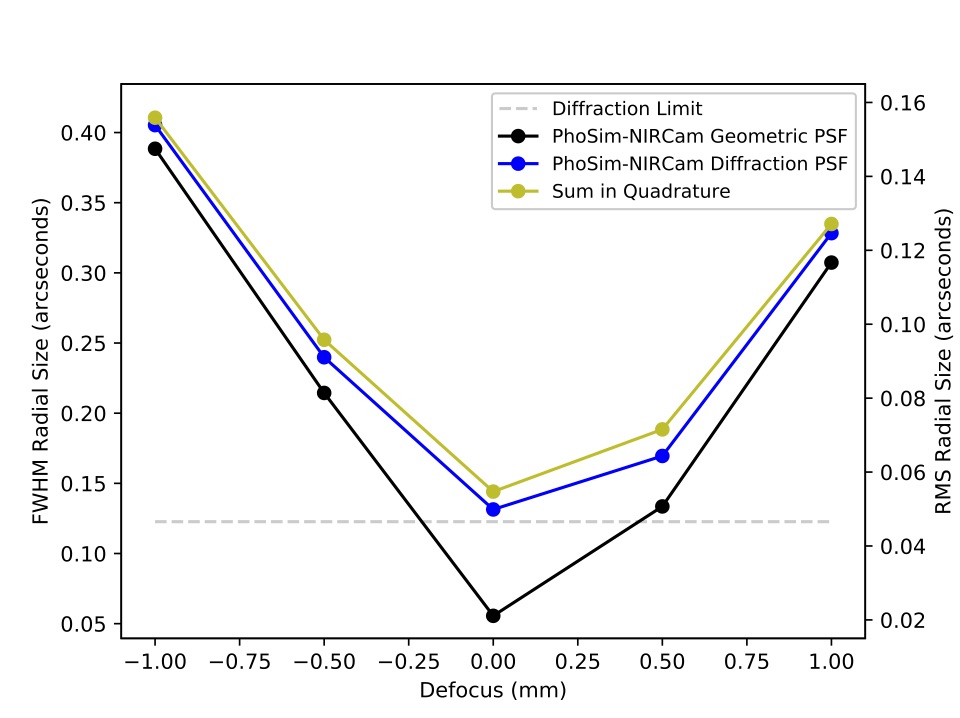}
\caption{\label{fig:psfsize1}
PhoSim PSF size versus defocus of image plane. The blue line is the total simulated geometric and diffraction PSF simulated by PhoSim propagation of geometric rays to the exit pupil with photon kicks from the FFT of the JWST tricontagon pupil geometry. The black line is the geometric-only component with a cylindrically-symmetric surface only. The yellow line is the sum of in quadrature of the geometric component (black line) and the diffraction limit (dashed gray line), which rightly approximates the blue line. Data were simulated at 3.56 $\mu$m near the center of the field of view (field point 5). The blue line is slightly underestimated due to the image cutout cutting off some of the PSF. Uncertainties are on the order of $10^{-8}$ arcseconds for the PhoSim data and $10^{-5}$ arcseconds for the WebbPSF data.}
\centering
\includegraphics[width=1.0\columnwidth]{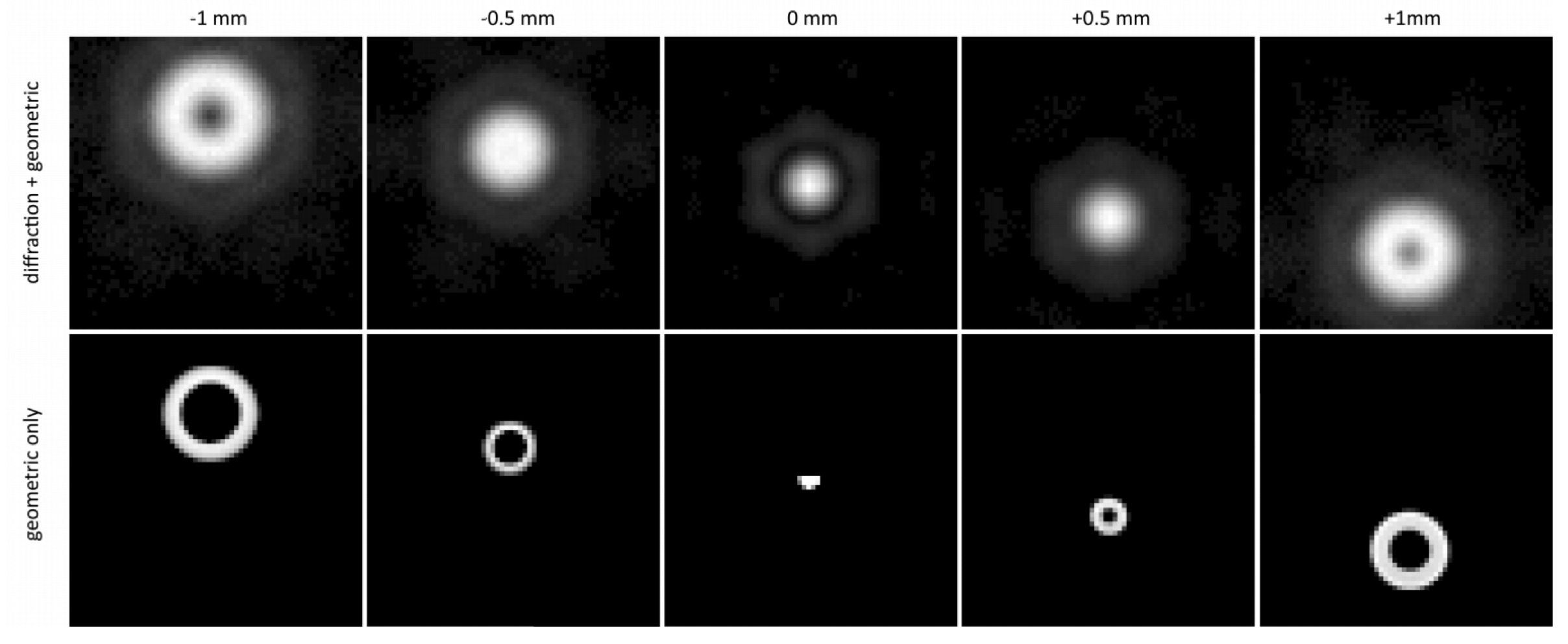}
\caption{\label{fig:psfsize2}
PhoSim PSF versus defocus of image plane. Data obtained at 3.56 $\mu$m near the center of the field of view (field point 5) using an oversampled image cutout. The top row shows simulated PSFs with diffraction and geometric physics on (Fig.~\ref{fig:psfsize1}, blue line). The bottom row shows simulated PSFs with the geometric raytrace approximation only (Fig.~\ref{fig:psfsize1}, black line). The ring shape is due to our cylindrically-symmetric approximation of the primary mirror geometry for the raytrace. Complete results for the geometric PSFs may be investigated in the future by coupling the tricontagon mirror geometry to the PhoSim raytrace code.}
\end{figure}

Fig.~\ref{fig:psfsize1} and \ref{fig:psfsize2} demonstrate how the PSF size varies as function of defocus of the image plane in PhoSim. We move the detector surface through the optimal focus and measure the size of the geometric component of the PSF, and the total PSF. The results show that PhoSim is approximating the total PSF in accordance with Eq.~\ref{eq:diff}.

\begin{figure}[ht]
\centering
\includegraphics[width=1\columnwidth]{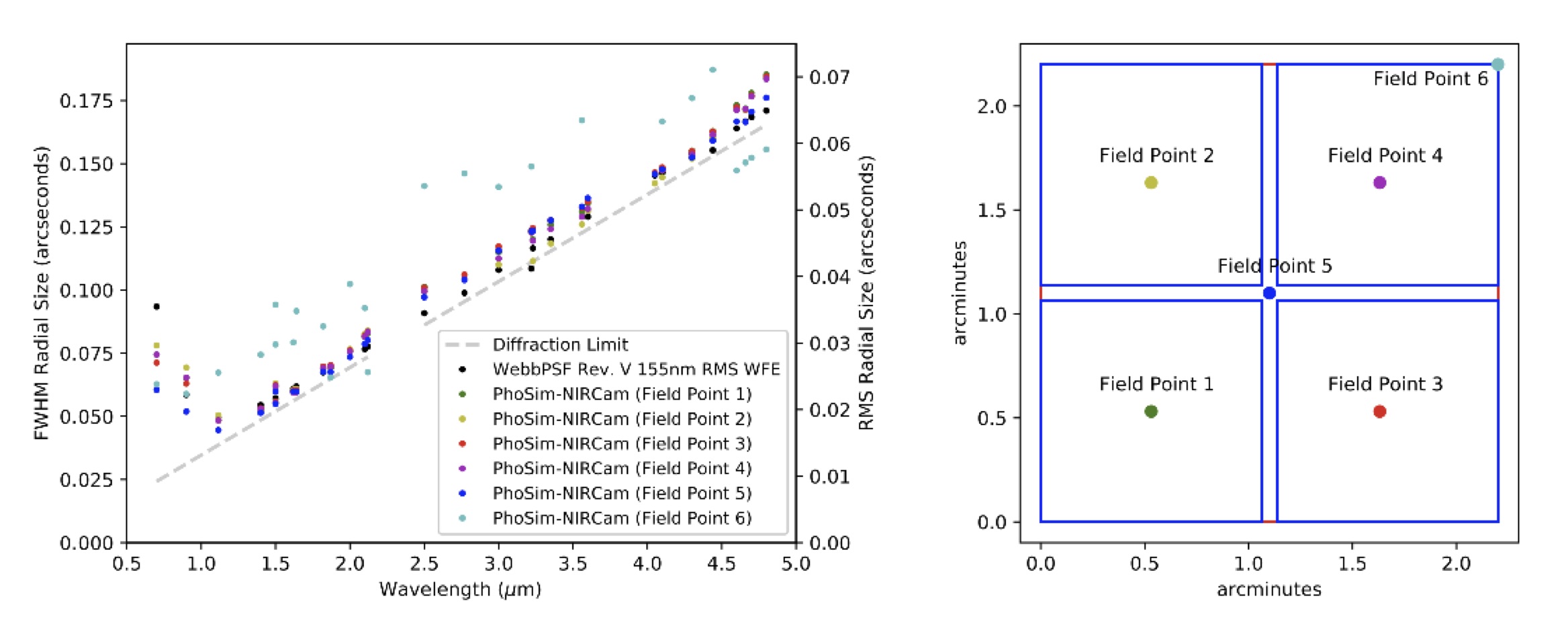}
(a) \hspace{5.7cm} (b)
\caption{\label{fig:psfsize3}
PSF radial size measured from the centroid of the PSF versus wavelength at $0$ mm defocus are shown in (a). Uncertainties are on the order of $10^{-8}$ arcseconds for the PhoSim data and $10^{-5}$ arcseconds for the WebbPSF data. Field point 5 is near the center of the field of view. Field points 1-4 are near the centers of chips 1-4. Field point 6 is near the edge of the field of view. We emphasize that this data may not represent the final PSFs, which depends on the final design configuration of the observatory and will need to be calibrated in-orbit. (b) shows the locations of the test field points. Also shown are the approximate outlines of the chips (blue squares = SW, enclosing red square = LW) in the PhoSim-NIRCam field of view. Note, this has yet to be matched to the final, exact chip positions.}
\end{figure}

Fig.~\ref{fig:psfsize3} shows the PSF radial size versus wavelength at each filter location for 5 different field positions. Data points for PhoSim results and WebbPSF\cite{Perren:2012,Perren:2014} revision V results (155 nm RMS WFE OPD model; OTE+NIRCam) are shown for comparison. These results are highly dependent on the final details of the optics and defocus settings. Nevertheless, the results appear consistent with WebbPSF. The large size of the PSF below 1.2 $\mu$m is due to the large geometric PSF component at those wavelengths. In our tests, different focus positions of the image plane exist which removes this, at the undesirable expense of slightly larger PSFs at wavelengths near or above 2 $\mu$m.

\begin{figure}[ht]
\centering
\includegraphics[width=0.95\columnwidth]{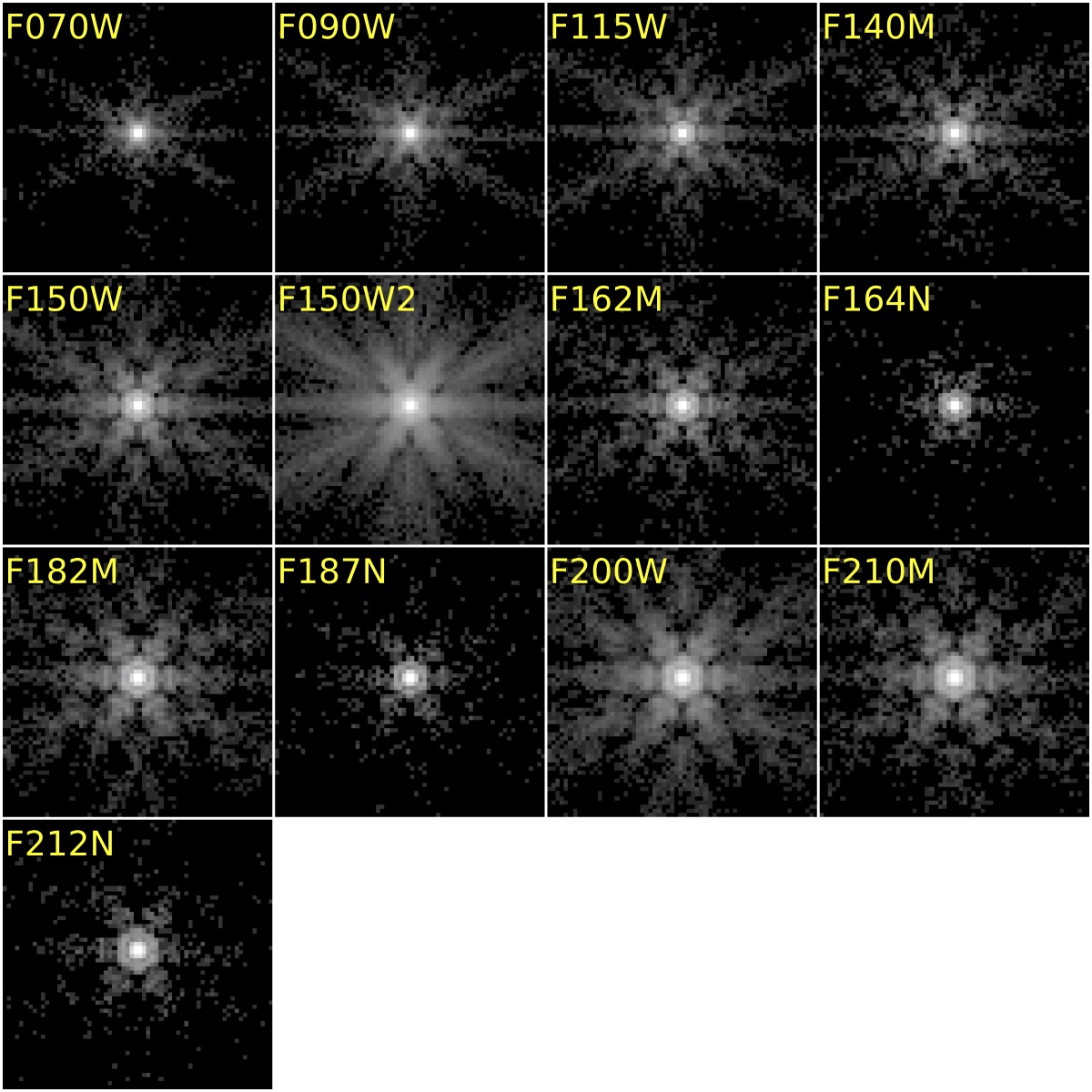}
\caption{\label{fig:psfs1}
PhoSim-NIRCam PSFs for each SW filter at $0$ mm defocus at field point 5. Near the center of the field of view, the PSFs are typically near diffraction-limited and well-behaved.}
\end{figure}

\begin{figure}[ht]
\centering
\includegraphics[width=0.95\columnwidth]{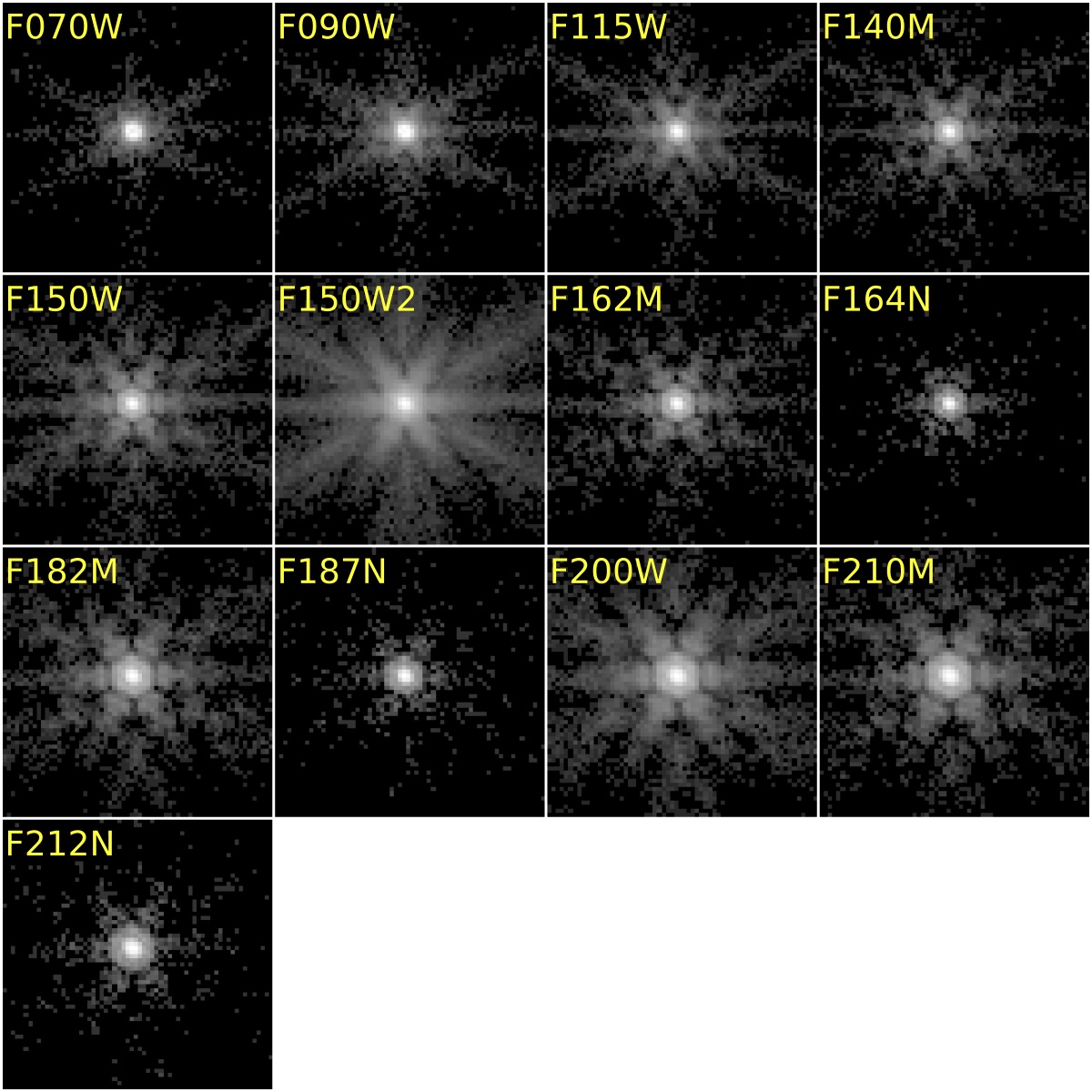}
\caption{\label{fig:psfs3}
PhoSim-NIRCam PSFs for each SW filter at $0$ mm defocus at field point 6. Near the edge of the field of view, the PSFs are typically not diffraction-limited and more distorted.}
\end{figure}

\begin{figure}[ht]
\centering
\includegraphics[width=0.95\columnwidth]{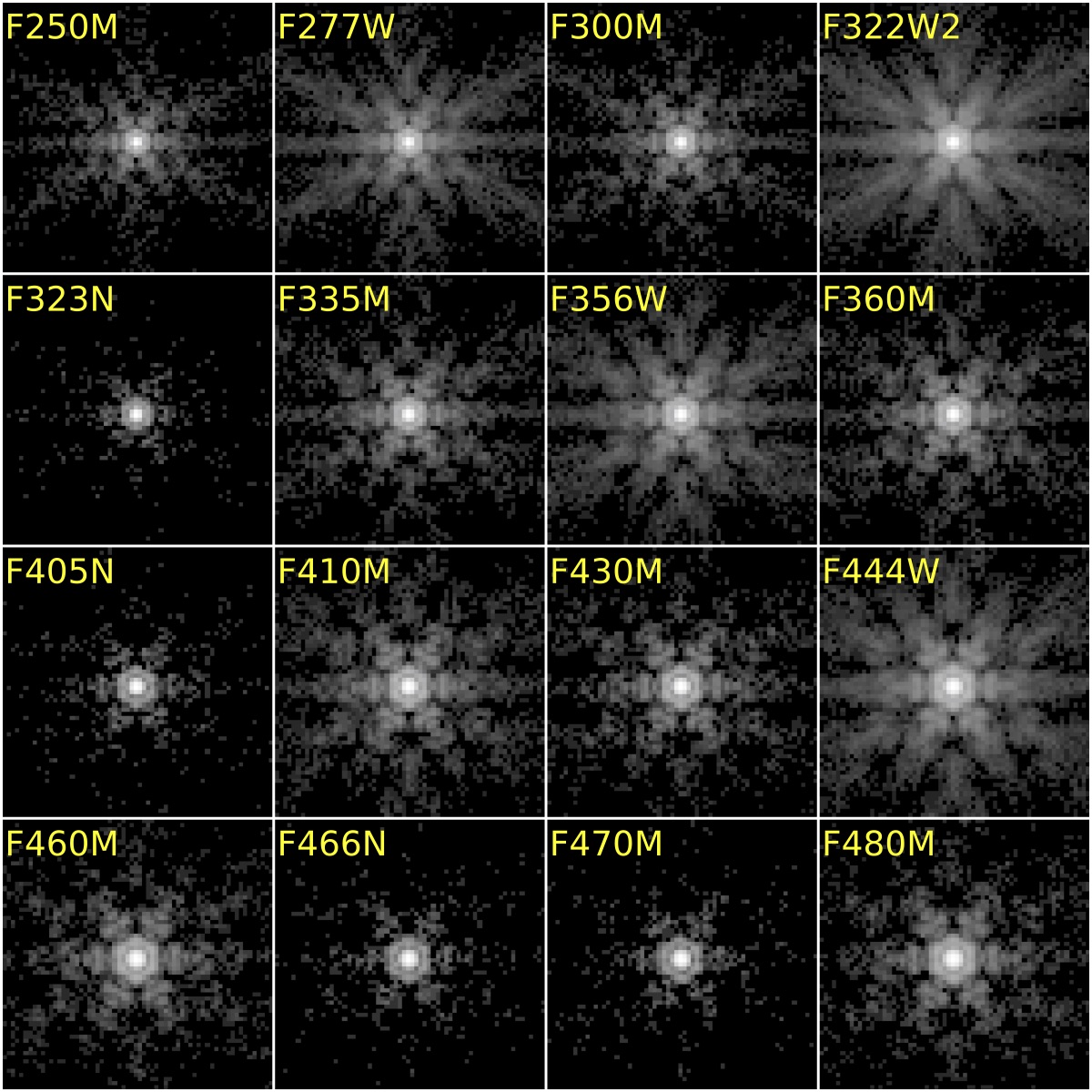}
\caption{\label{fig:psfs2}
PhoSim-NIRCam PSFs for each LW filter at $0$ mm defocus at field point 5. Near the center of the field of view, the PSFs are typically near diffraction-limited and well-behaved.}
\end{figure}

\begin{figure}[ht]
\centering
\includegraphics[width=0.95\columnwidth]{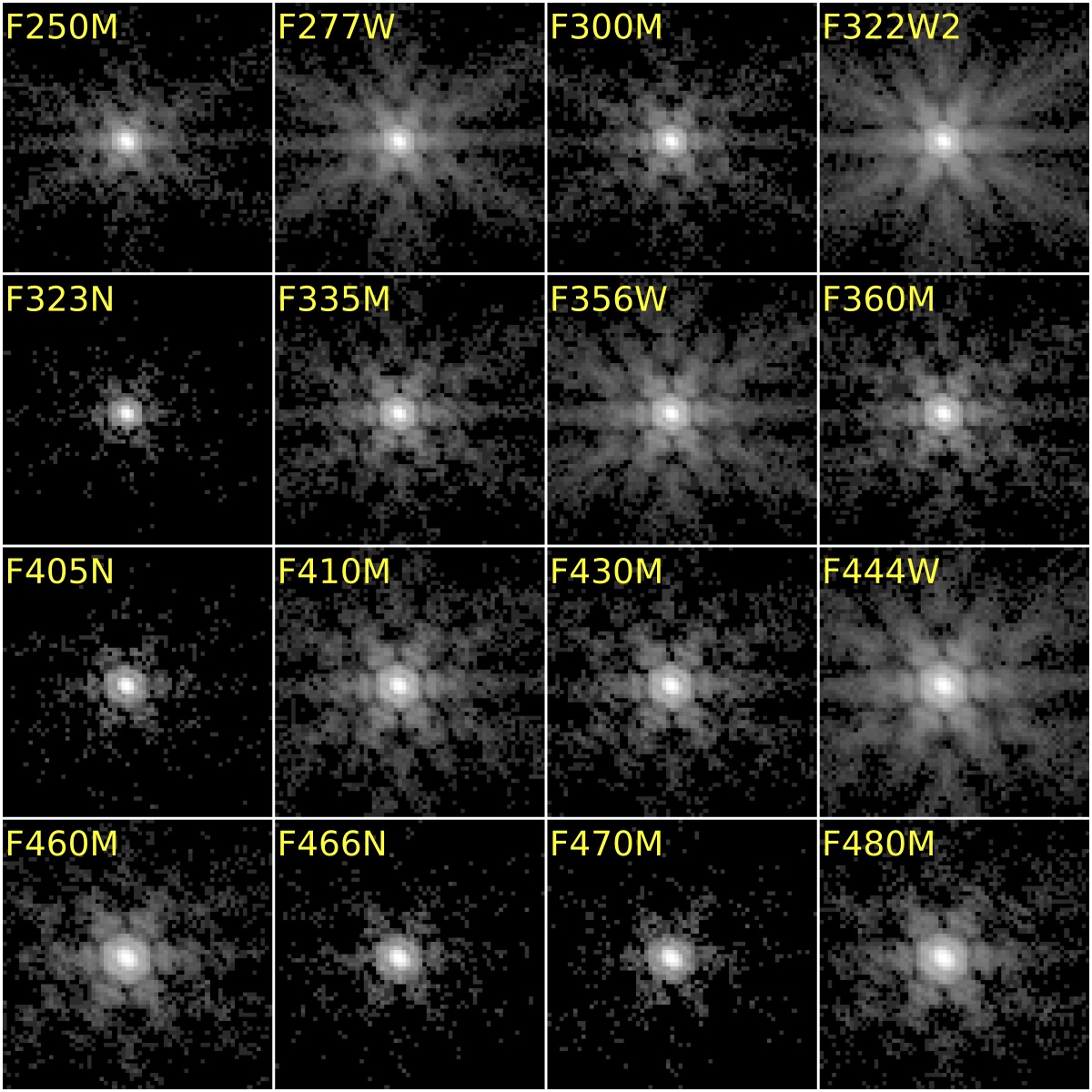}
\caption{\label{fig:psfs4}
PhoSim-NIRCam PSFs for each LW filter at $0$ mm defocus at field point 6. Near the edge of the field of view, the PSFs are typically not diffraction-limited and more distorted.}
\end{figure}

The results shown here are consistent with WebbPSF results and the requirement that NIRCam images be near diffraction-limited above 2 $\mu$m. Although the exact results are somewhat dependent on the SED of choice, especially for the wide-band filters, and the final optical configuration. Fig.~\ref{fig:psfs1}-\ref{fig:psfs4} show a sample of these PSFs for field points 5 and 6. The figures show that the PSFs are near diffraction-limited and well-behaved around the field center, but are noticeably distorted toward the field edge.

\cleardoublepage

\subsection{Simulation of Extragalactic Sky}

To demonstrate the capabilities of PhoSim-NIRCam, we have simulated an extragalactic blank-sky field using the source catalog created from a real space near-infrared image taken with the WFC3-IR camera on the Hubble Space Telescope (HST).\footnote{\footnotesize The simulated images as well as the code and input catalogs used for this simulation are available at \url{https://fenrir.as.arizona.edu/phosim}.}

\begin{figure}[ht]
\centering
\includegraphics[width=0.95\columnwidth]{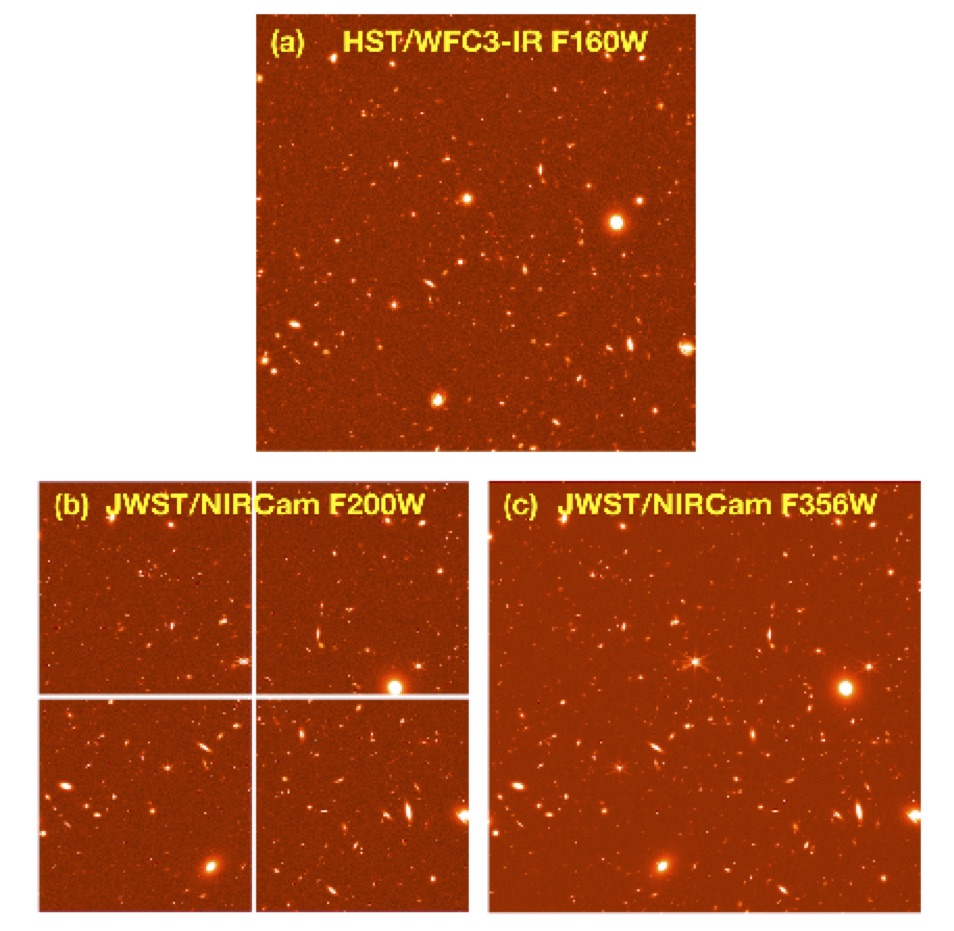}
\\
\caption{\label{fig:sim}
Comparison of a real HST/WFC3-IR F160W image (top) and simulated JWST/NIRCam F200W (bottom left) and F356W (bottom right) images created by PhoSim-NIRCam.  The filter numbers refer to wavelength in $\mu$m (e.g., the effective wavelength of F160W is 1.6 $\mu$m).  The input source catalog for the simulation was produced from the HST image, and includes morphological information based on S\'ersic 2D models.  A flat-$f_\nu$  SED was used to extrapolate source brightnesses to 2 and 3.6 $\mu$m, and a sky background of 0.1 MJy/sr was included.   The NIRCam SW F200W image consists of four quadrants of H2RG detectors (0.03"/pixel) while the LW F356W image is produced by one H2RG detector (0.06"/pixel).  The figure clearly demonstrates that PhoSim-NIRCam is capable of producing realistic JWST/NIRCam images for both the SW and LW channels.  Note the diffraction spikes seen for bright sources in the NIRCam images produced by the hexagonal primary mirror of JWST.
}
\end{figure}

Fig.~\ref{fig:sim} shows the comparison between the original WFC3-IR F160W image (the top panel; part of the CANDELS GOODS-S data\cite{Grogin2011,Koekemoer2011}) and simulated NIRCam SW F200W image (the bottom left panel) and LW F356W image (the bottom right panel).  Although the JWST primary mirror ($D = 6.5$ m) is significantly larger than that of the HST ($D = 2.4$ m), the observed wavelengths of these simulated NIRCam images are also longer (especially with the F356W image), resulting in comparable image sizes.

The figure clearly demonstrates that PhoSim-NIRCam is capable of producing realistic JWST/NIRCam images for both the SW and LW channels, including the diffraction spikes produced by the hexagonal primary mirror of the JWST. Although an integration time of 4 hours was assumed to produce these NIRCam images, the image depth here is essentially limited by the CANDELS HST source catalog used for the simulation.  Actual 4-hour NIRCam images would look much deeper with many more fainter objects.  Here, the HST-produced source catalog was used to allow direct comparison between the real and simulated images, but to simulate more realistic deeper NIRCam images with a variety of galaxy SEDs, the use of a properly constructed mock catalog\cite{Williams:2018} would be more appropriate (such a simulation is currently underway).

\section{Conclusions, Limitations, and Future Work}

We harness the power of PhoSim, demonstrating the capability to simulate high-fidelity NIRCam images from a realistic catalog of stars and galaxies. Our end-to-end physics-based method simulates one photon at a time to replicate the relevant effects on NIRCam PSF morphology and overall image characteristics relevant for background-limited observations.\footnote{\footnotesize The PhoSim-NIRCam source code and ISC files describing the entire PhoSim-NIRCam model are open-source and available at \url{https://bitbucket.org/phosim/phosim_release}.} Software updates including additional features and bug fixes are published on a regular basis.

PhoSim-NIRCam may be applied to better understand systematics in NIRCam images, especially galaxy morphology, weak lensing, and morphology of other extended sources. The initial avenues for applying PhoSim-NIRCam are the GTO and ERS programs, and in-orbit commissioning.

In this work, we present a method to simulate space-based optical and infrared instruments within the comprehensive PhoSim framework. We show initial PhoSim-NIRCam results that approximate the wavelength- and field- dependent PSF behavior. We demonstrate PhoSim-NIRCam's capability to simulate NIRCam PSFs with various physics independently, although we do not yet claim our results are an accurate model of the instrument's ultimate performance. In the future, a more complete validation and modeling campaign will be performed in conjunction with the JWST in-orbit commissioning.

Planned extensions of this work include implementing the proper spatial pattern and power spectrum of NIRCam readout noise from pyNRC\cite{Leisenring2018} along with a realistic FITS file format for use with JWST image pipeline (e.g. header keywords and naming scheme). We also plan to include a more realistic cosmic ray signature and rates for space telescopes. It may also be interesting to investigate a more realistic thermal model of the telescope and detectors, extending recent work coupling photon Monte Carlo methods to opto-mechanical deformation of ground-based optics\cite{Peterson2019}.

Finally, a model to couple the geometric raytrace to the primary mirror tricontagon geometry, surface perturbations, and figure errors that properly affect the ellipticity and higher-order spatial content of the geometric PSF will be investigated in the future. This work will be essential to matching the final performance of NIRCam after launch.

We welcome greater involvement from the JWST community to supplement these efforts. Extensions to other space-based or ground-based telescopes would also be straightforward following the methods developed in this work.

\acknowledgments

JRP \& CJB acknowledge funding from Purdue University and NASA through Subcontract \#366623 from NASA Contract NAS50210 with the University of
Arizona. We thank Dr. Christina Williams for providing the CANDELS GOODS-S source catalog for our image simulation. We acknowledge use of Matplotlib\cite{Hunter:2007}, a community-developed Python library for plotting. This research has made use of NASA's Astrophysics Data System.
\small
\begin{multicols}{2}
\bibliography{report}   
\bibliographystyle{spiejour}   
\end{multicols}

\end{spacing}
\end{document}